\documentclass[10pt,aps,prl,twocolumn,superscriptaddress]{revtex4}

\usepackage[utf8]{inputenc}
\usepackage{hyperref}
\usepackage{graphicx}
\usepackage{amsmath,amssymb}
\usepackage{color}
\usepackage{subfigure}
\usepackage{url}

\graphicspath{{Images/}}

\begin{document}

\title{High speed optical holography of retinal blood flow
}

\author{M. Pellizzari}
\affiliation{
Centre National de la Recherche Scientifique (CNRS) UMR 7587, Institut Langevin. Paris Sciences et Lettres (PSL) Research University. Fondation Pierre-Gilles de Gennes, Institut National de la Sant\'e et de la Recherche M\'edicale (INSERM) U 979, Universit\'e Pierre et Marie Curie (UPMC), Universit\'e Paris 7. \'Ecole Sup\'erieure de Physique et de Chimie Industrielles ESPCI Paris - 1 rue Jussieu. 75005 Paris. France\\
}

\author{M. Simonutti}
\affiliation{
Institut de la Vision, INSERM UMR-S 968. CNRS UMR 7210. UPMC. 17 rue Moreau, 75012 Paris. France
}

\author{J. Degardin}
\affiliation{
Institut de la Vision, INSERM UMR-S 968. CNRS UMR 7210. UPMC. 17 rue Moreau, 75012 Paris. France
}

\author{J.-A. Sahel}
\affiliation{
Institut de la Vision, INSERM UMR-S 968. CNRS UMR 7210. UPMC. 17 rue Moreau, 75012 Paris. France
}

\author{M. Fink}
\affiliation{
Centre National de la Recherche Scientifique (CNRS) UMR 7587, Institut Langevin. Paris Sciences et Lettres (PSL) Research University. Fondation Pierre-Gilles de Gennes, Institut National de la Sant\'e et de la Recherche M\'edicale (INSERM) U 979, Universit\'e Pierre et Marie Curie (UPMC), Universit\'e Paris 7. \'Ecole Sup\'erieure de Physique et de Chimie Industrielles ESPCI Paris - 1 rue Jussieu. 75005 Paris. France\\
}

\author{M. Paques}
\affiliation{
Institut de la Vision, INSERM UMR-S 968. CNRS UMR 7210. UPMC. 17 rue Moreau, 75012 Paris. France
}

\author{M. Atlan}
\affiliation{
Centre National de la Recherche Scientifique (CNRS) UMR 7587, Institut Langevin. Paris Sciences et Lettres (PSL) Research University. Fondation Pierre-Gilles de Gennes, Institut National de la Sant\'e et de la Recherche M\'edicale (INSERM) U 979, Universit\'e Pierre et Marie Curie (UPMC), Universit\'e Paris 7. \'Ecole Sup\'erieure de Physique et de Chimie Industrielles ESPCI Paris - 1 rue Jussieu. 75005 Paris. France\\
}

\date{\today}

\begin{abstract}

We performed non-invasive video imaging of retinal blood flow in a pigmented rat by holographic interferometry of near-infrared laser light backscattered by retinal tissue, beating against an off-axis reference beam sampled at a frame rate of 39 kHz with a high throughput camera. Local Doppler contrasts emerged from the envelopes of short-time Fourier transforms and the phase of autocorrelation functions of holograms rendered by Fresnel transformation. This approach permitted imaging of blood flow in large retinal vessels ($\sim \,$ 30 microns diameter) over 400 $\times$ 400 pixels with a spatial resolution of $\sim \,$ 8 microns and a temporal resolution of $\sim \,$ 6.5 ms.

\end{abstract}

\maketitle

Retinal blood flow plays a central role in the pathophysiology of many eye diseases~\cite{PradaHarrisGuidoboni2015}, either through capillary occlusion, panretinal hypoperfusion or uneveness of flow distribution. Noninvasive analysis of retinal hemodynamics has been achieved through various coherent light schemes including standard laser Doppler velocimetry~\cite{Riva1972, RivaGeiserPetrig2009, MentekTruffer2015}, spatial~\cite{ChengDuong2007,SriencKurthNelson2010,PonticorvoCardenas2013} and temporal~\cite{Tamaki1994, Sugiyama2010, SaitoSaito2015} speckle contrast analysis, stroboscopic fundus cameras~\cite{IzhakyNelson2009}, functional optical coherence tomography~\cite{Wang2010, DziennisQinShi2015} and its full-field variant~\cite{SpahrHillmann2015}. Each of these approaches has its own limits in terms of sensitivity, spatial and/or temporal resolution, or lateral field of view. Holographic imaging of the retina was demonstrated with photographic plates~\cite{Calkins1970, Wiggins1972, OhzuKawara1979, Tokuda1980}; since the advent of digital cameras, Doppler imaging of blood flow can also be achieved by frequency-tunable time-averaged holographic interferometry at very low irradiance levels~\cite{SimonuttiPaquesSahel2010, MagnainCastelBoucneau2014}. But the major drawback of this approach is the need for sequential frequency tuning of the Doppler components, which hinders high temporal resolution imaging. The results reported here overcome this limitation : we performed holographic interferometry with a high throughput camera for bidimensional mapping of retinal flow velocity, and its local pulsatile component.


%
\begin{figure}[]
\centering
\includegraphics[width = 8 cm]{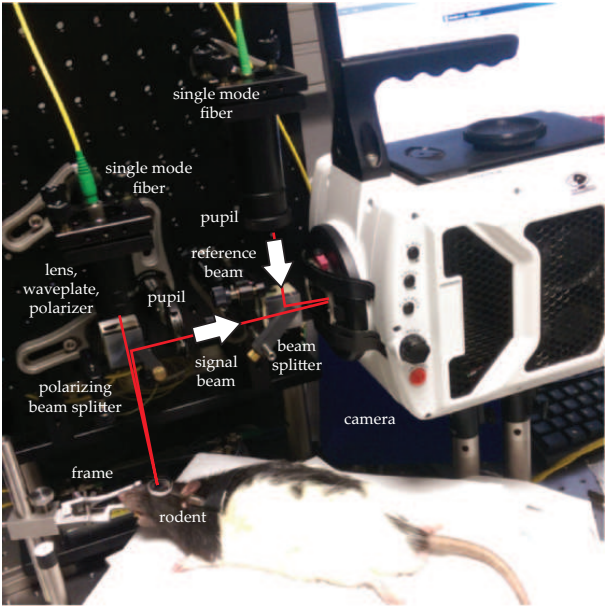}
\includegraphics[width = 8 cm]{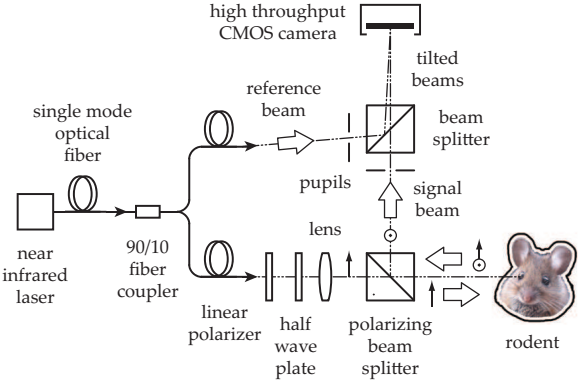}
\caption{Picture (top) and sketch (bottom) of the fibered Mach-Zehnder holographic interferometer used for retinal imaging. Off-axis interferograms of near-infrared radiation are recorded by the sensor array of a high speed camera.}
\label{fig_Setup}
\end{figure} 

The experimental imaging scheme designed for this study is sketched in Fig.~\ref{fig_Setup}; it consists of a fibered Mach-Zehnder optical interferometer in off-axis configuration. The optical source used for the experiments was a $\sim$ 45 mW, single-mode, fibered diode laser (Newport SWL-7513-H-P) at wavelength $\lambda = 785$ nm, and optical frequency $\omega_{\rm L} / (2 \pi) = 3.8 \times 10^{14} \, \rm Hz$. Two adult rats were used for the preparations (Elevage Janvier, Saint-ile-Le Genest, France). Anesthesia was induced by intraperitoneal injection of 40 mg/kg ketamine (Virbac) and 0.14 mg/kg domitor (Vetoquinol). Topical tropicamide (Cibavision) was administered for pupil dilation. Each animal was placed on its side under the illumination beam. The head was supported so that the iris was perpendicular to the illumination axis. After administration of one drop of 1.6 mg/0.4 ml topical oxybuprocaine (Théa), a coverslip was applied on a ring surrounding the globe, with Lubrithal (Dechra) as contact medium. A heating plate maintained body temperature at 37 $^\circ$C. Experiments were performed in conformity with the European Community Council Directive of 22nd September 2010 (010/63/UE) and approved by the local ethics committee (C2EA -59, 'Paris Centre et Sud', authorization number: 2012-0068). Retinas were illuminated with a continuous-wave laser beam of $\sim$ 1.6 mW power, over $\sim 3 \, {\rm mm} \times 3 \, {\rm mm}$. This irradiation level is compliant with the guidelines for exposure of the eye to optical radiation from ocular instruments~\cite{SlineyAronRosa2005} and the International Organization for Standardization norm ISO 15004-2:2007. The power of the reference wave (local oscillator, LO) impinging over the full sensor was $\sim 400 \, \mu {\rm W}$. In the object arm, a polarizing beam splitter cube was used to illuminate the preparation under linearly polarized light and collect the cross-polarized backscattered component, in order to increase the relative weight of multiply scattered Doppler-shifted photons with respect to photons scattered once \cite{Schmitt1992}, and benefit from the birefringence properties of blood vessels~\cite{WeberChenev2004}. The backscattered optical field $E$ was mixed with the LO field $E_{\rm LO}$ with a non-polarizing beam splitter cube, tilted by $\sim 1 ^\circ$ to ensure off-axis recording conditions. Light-tissue interaction resulted in a local phase variation $\phi(t)$ of the backscattered laser optical field $E(t) = {\cal E}(t)  \exp \left[ i \omega_{\rm L} t + i \phi(t) \right]$, which was mixed with the LO field from the reference channel $E_{\rm LO}(t) = {\cal E}_{\rm LO}(t) \exp \left[ i \omega_{\rm L}t + i \phi_{\rm LO}(t)\right]$. The quantity $i$ is the imaginary unit, while ${\cal E}_{\rm LO}(t)$, ${\cal E}(t)$ are the envelopes, and $\exp[i\phi_{\rm LO}(t)]$, $\exp[i\phi(t)]$ are the phase factors of the fields, respectively. In this description, we consider a reference wave devoid of temporal fluctuations in amplitude and phase in the (filtered) detection bandwidth, so we can write $E_{\rm LO}(t) = {\cal E}_{\rm LO} \exp \left[ i \omega_{\rm L}t \right]$, where ${\cal E}_{\rm LO}$ is constant in time. Optical interferograms of $768 \times 768$ pixels of coordinates $(x,y)$ were digitally acquired by the sensor array of a high throughput camera (Ametek - Phantom V2511), at a frame rate of $\tau_{\rm S}^{-1} = \omega_{\rm S} / (2 \pi) = 39.0 \, \rm kHz$, with a frame exposure time of $\tau_{\rm E} = 25.0 \, \mu {\rm s}$, and a pixel size $d_{\rm px} = 28 \, \mu \rm m$. The distance between the eye and the sensor was $L\sim 30 \, \rm cm$. The cross-beating component $H = {\cal E}_{\rm LO}^* {\cal E} \exp \left( i \phi \right)$ of the interferogram $I = \left| {\cal E} \right|^2 + \left| {\cal E}_{\rm LO} \right|^2 + H + H^*$, where $^*$ denotes the complex conjugate, was filtered spatially~\cite{Cuche2000} from the other interferometric contributions.

\begin{figure}[]
\centering
\includegraphics[width = 7.3 cm]{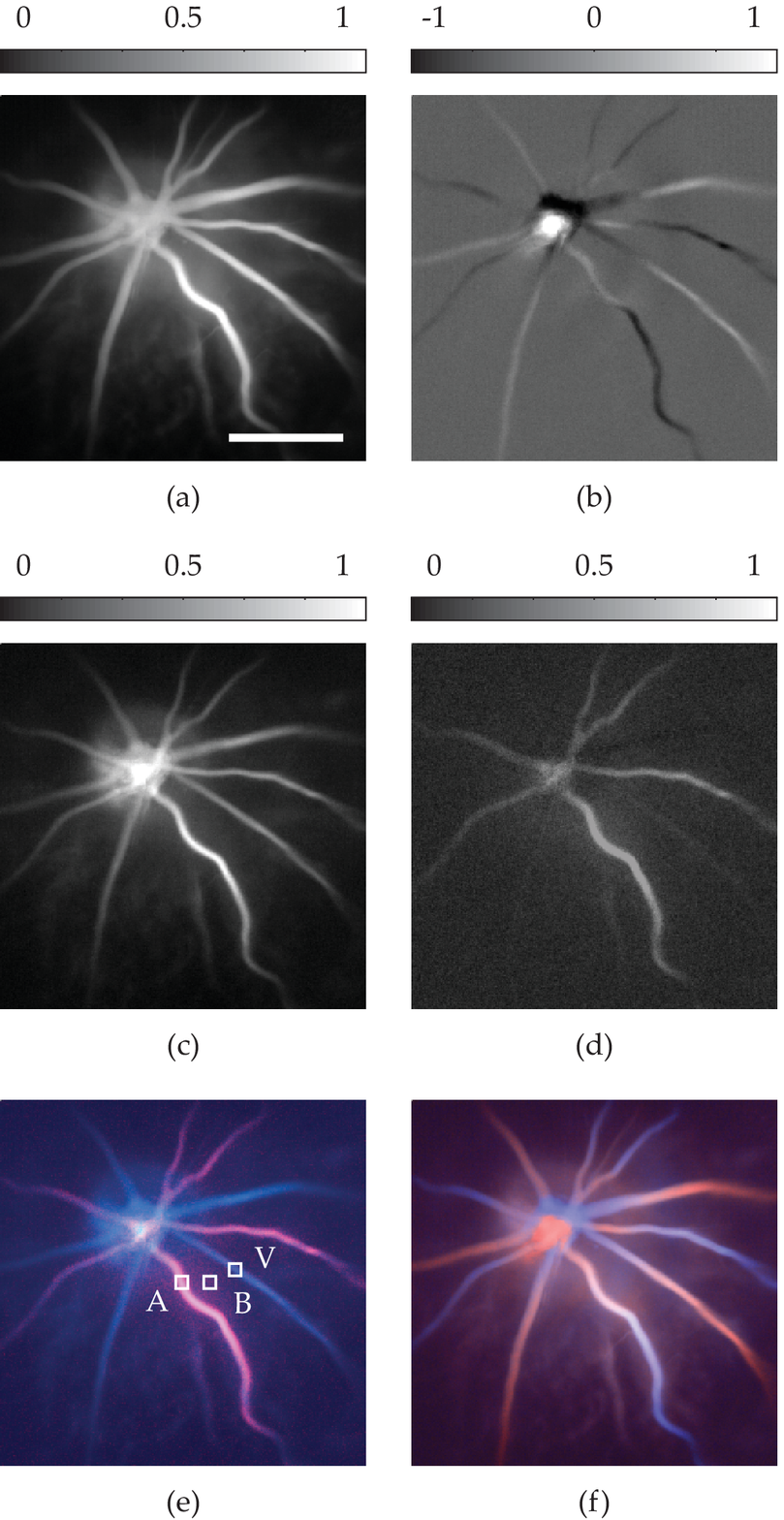}
\caption{Zeroth (a), first (b), and second (c) moments $\left< \omega^n \right>$, $n= 0, 1, 2$, of the envelope $S$ (Eq.~\ref{eq_Moments}); \href{https://youtu.be/iKcfCe04ffw}{Visualization 1}. Zero-lag correlation map $\int \left< \omega^2 \right>_A(t) \times \left< \omega^2 \right>(x,y,t) {\rm d}t$ with the arterial signal $\left< \omega^2 \right>_A$ (d). Composite color images of the zeroth and second moment [(e) and \href{https://youtu.be/skjNnhSI35Y}{Visualization 2}], and zeroth and first moment [(f) and \href{https://youtu.be/35-iVY-_JnQ}{Visualization 3}]. Scale bar $\sim$ 1 mm. Color scales : arbitrary units (a.u.).
}\label{fig_Images}
\end{figure}
\begin{figure}[]
\centering
\includegraphics[width = 7.3 cm]{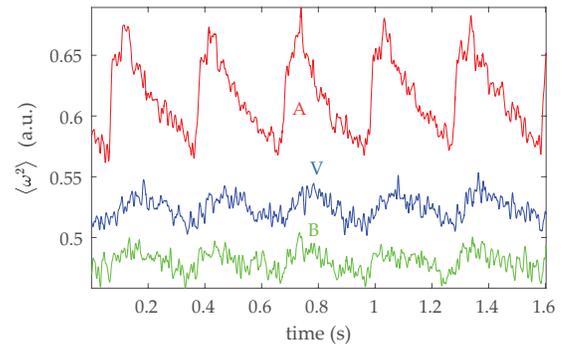} 
\caption{Second moment moment of the distribution $S$ (Eq.~\ref{eq_Moments}) versus time, averaged spatially in the regions of interest A, V, B, depicted in Fig.~\ref{fig_Images}.
}\label{fig_MomentsVsTime}
\end{figure}
%


The image rendering of off-axis, complex-valued holograms $H(x,y,t)$ was performed by discrete Fresnel transformation~\cite{PiedrahitaQuinteroCastanedaGarciaSucerquia2015} of recorded interferograms $I(x,y,t)$
\begin{eqnarray}\label{eq_FresnelTransform}
\nonumber H(x,y,t) = \frac{i}{\lambda z}\exp \left( -ikz \right) \iint I(x',y',t)\\
\times \exp \left[\frac{-i \pi}{\lambda z} \left((x-x')^2 + (y-y')^2\right) \right] {\rm d}x' {\rm d}y'
\end{eqnarray}
where $k = 2\pi/\lambda$ is the optical wave number, and $z = 0.91 \, \rm m$ is the hologram reconstruction distance. The discrepancy between the numerical reconstruction distance $z$ and the physical distance $L$ between the retina and the camera sensor is explained by the non-zero curvature of the LO field in the sensor plane. The calculation grid $(x',y')$ was zero-padded to a $1024 \times 1024$ pixels array. The lateral field of view of the reconstructed hologram~\cite{Siegman1966} is about $\lambda L / d_{\rm px} = 8.4 \, \rm mm$, from which the image covers a fraction $400/1024$. Neglecting the refraction of the eye's lens gives a field of view of the reported retinal images of $\sim 3.3\, {\rm mm} \times 3.3 \, {\rm mm}$, and a spatial resolution of $\sim 8 \, \mu {\rm m}$.\\


Signal processing was performed on a series of $2^{16}$ consecutive holograms $H$ by the following steps: First, the stack of complex-valued holograms $H$ was processed temporally by a 2 kHz cutoff high-pass 3rd-order Butterworth filter~\cite{Butterworth1930}, to remove low-frequency components. This operation could have been avoided, but doing so led to moment maps (Eq.~\ref{eq_Moments}) of better quality than without filtering. Then, the envelope of the short-time Fourier transform of $H$ was formed
\begin{equation}\label{eq_STFT}
S(x,y,t,\omega) = \left| \int H(x,y,\tau) g(t-\tau) e^{-i \omega \tau} \, {\rm d}\tau \right|^2 
\end{equation}
where $g(t)=\exp(-t^2/(2\tau_0^2))$ is an apodization window of width $2 \tau_0 \simeq 6.5 \, \rm ms$. Assessing blood perfusion indicators from the first moments of the power spectrum distribution $S$ of the detector signal is commonplace for laser Doppler sensors~\cite{BonnerNossal1981, Leutenegger2011}; here we formed the three first moments ($n=0,1,2$) in
\begin{equation}\label{eq_Moments}
\left< \omega^n \right> (x,y,t) = \displaystyle \int S(x,y,t,\omega) \omega^n {\rm d} \omega 
\end{equation}
reported in Fig.~\ref{fig_Images}(a) ($n=0$), Fig.~\ref{fig_Images}(b) ($n=1$), Fig.~\ref{fig_Images}(c) ($n=2$). In these images, the three first moments $\left< \omega^n \right>$ were calculated with a time step of $\sim$ 2.5 ms, but the actual temporal resolution is limited by the width $2\tau_0$ of the apodization window. On the zeroth moment $\left< \omega^0 \right>$, the set of six retinal arteries and veins is revealed. The first moment $\left< \omega^1 \right>$ characterizes the centroid of the spectral envelope in the Shannon bandwidth of the measurement $[-19.5 \,{\rm kHz}, 19.5 \,{\rm kHz}[$, and appears to be sensitive to the direction of the flow with respect to the optical axis. Cardiac cycles of $\sim 3\, \rm Hz$ are clearly revealed in the second moment $\left< \omega^2 \right>$, averaged in the regions "A", "V", and "B", and plotted versus time in Fig.~\ref{fig_MomentsVsTime}. In these results, time lags are observed between the peaks. The peak flow in a selected vein (V) occurs $\sim 25 \, \rm ms$ after the peak flow of an artery (A), and the peak of the background signal (B) appears in between. A time-dependent map of the arteries was formed by calculating $\int \left< \omega^2 \right>_A(t) \times \left< \omega^2 \right>(x,y,t) {\rm d}t$ at each position $(x,y)$, over a temporal window of 0.38 s (slightly more than one cardiac cycle). This zero-lag cross-correlation map, reported in Fig.~\ref{fig_Images}(d), enables clear distinction of arteries from veins. It is used as red channel of a composite color image and video, and mixed with $\left< \omega^0 \right>$ in the gray channel, and $\left< \omega^2 \right>$ in the blue channel. This composite map is reported in Fig.~\ref{fig_Images}(e). Another composite map is reported in Fig.~\ref{fig_Images}(f); it mixes the positive values of $\left< \omega^1 \right>$ in the red channel, the negative values of $\left< \omega^1 \right>$ in the blue channel, and $\left< \omega^0 \right>$ in the gray channel.

\begin{figure}[]
\centering
\includegraphics[width = 7.3 cm]{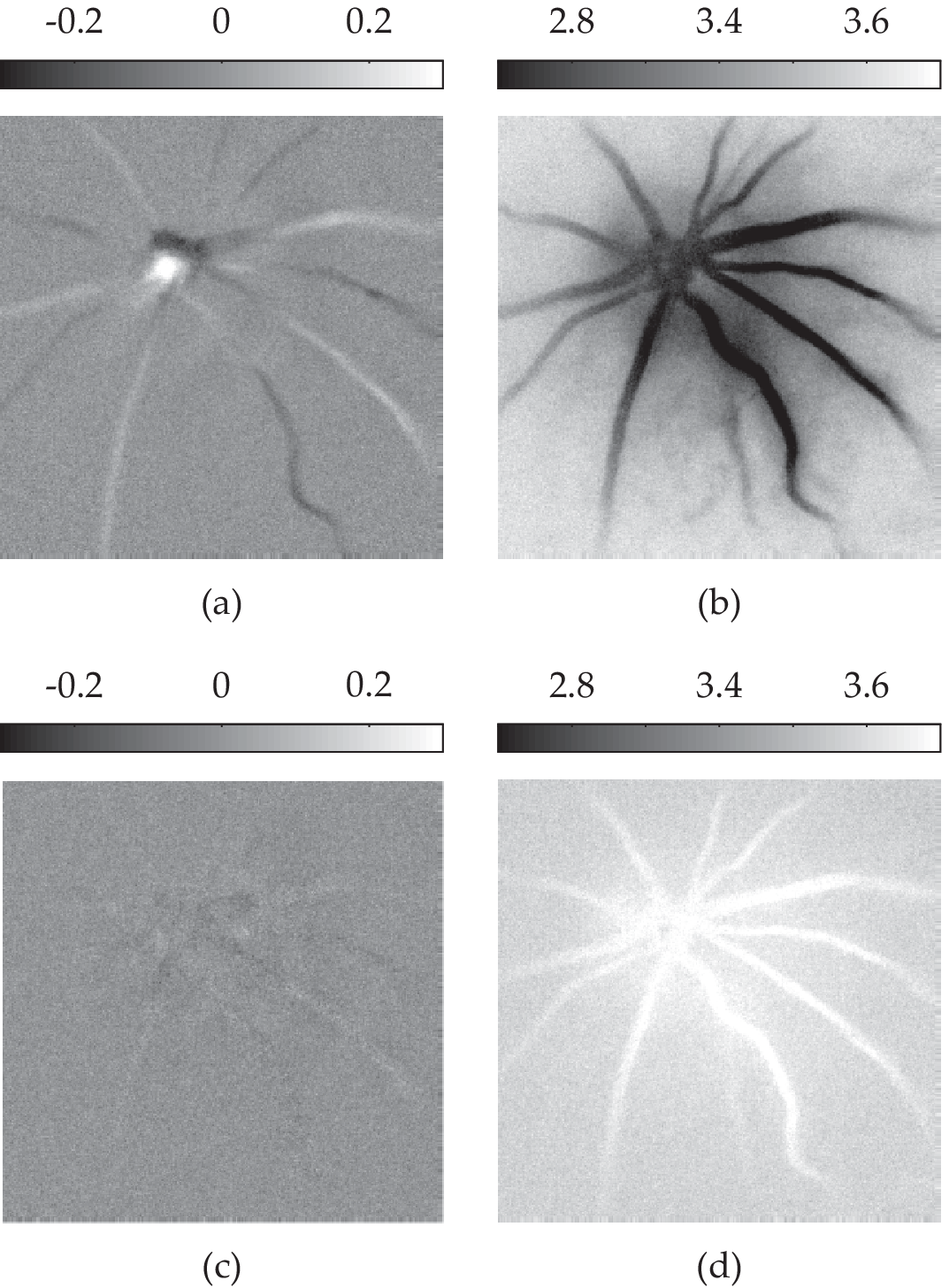}
\caption{Moments $\left< \Delta \phi_m^n \right>$ (Eq.~\ref{eq_PhaseMoments}) of the phase shift $\Delta \phi_m(t)$ for time lag $m\tau_{\rm S}$ (\href{https://youtu.be/2PYjum9e_Kw}{Visualization 4}). First ($n=1$ a,c) and second ($n=2$ b,d) moments for time lags $\tau_{\rm S}$ ($m=1$ a,b) and $4\tau_{\rm S}$ ($m=4$ c,d). Units: rad (a,c), rad$^2$ (b,d).
}\label{fig_PhaseImages}
\end{figure}
\begin{figure}[]
\centering
\includegraphics[width = 7.3 cm]{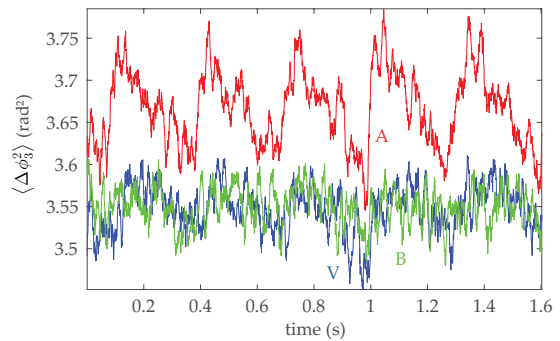} 
\caption{Second moment of the phase increase for a lag time $3\tau_{\rm S}$ versus time (Eq.~\ref{eq_PhaseMoments}), averaged spatially in the regions of interest A, V, B, depicted in Fig.~\ref{fig_Images}.
}\label{fig_MomentsPhaseVsTime}
\end{figure}

Additionally, we formed the product of complex-valued holograms $H(x,y,t)$ by their $m$-frame-lagged conjugates $H^*(x,y,t-m\tau_{\rm S})$, in order to seek temporal correlations, as performed in ultrasound imaging~\cite{Kasai1985}. Local phase shifts $\Delta \phi_m(t) = \phi(t) - \phi(t-m\tau_{\rm S})$, were assessed by calculating $\arg \left[ H(t)H^*(t-m\tau_{\rm S}) \right]$. Their $n$-th order moments 
\begin{equation}\label{eq_PhaseMoments}
\left< \Delta \phi_m^n \right> (x,y,t) = \tau_1^{-1}\displaystyle \int\limits_{t-\tau_1}^{t} \Delta\phi_m^n(x,y,\tau) \, {\rm d}\tau  
\end{equation}
were used to reveal local dynamics. The first and second moments of the phase shift $\Delta \phi_m$, measured with a gate $\tau_1 = 6.5 \, \rm ms$, for time lags $\tau_{\rm S}\simeq 25\, \mu \rm s$ ($m=1$) and $4\tau_{\rm S}\simeq 100\, \mu \rm s$ ($m=4$) are reported in Fig.~\ref{fig_PhaseImages}. In Fig.~\ref{fig_Images}(b) and Fig.~\ref{fig_PhaseImages}(a), positive Doppler frequency and phase shifts are observed in the central retinal artery and negative shifts occur in the central retinal vein. This result can be explained by the fact that the central artery and vein are parallel to the optic nerve, and roughly coaxial to the incident optical beam, and their flows are in opposite directions. For a lag time $3\tau_{\rm S}$, cardiac cycles are distinguished clearly in the second moment of the phase increase, as it can be seen in Fig.~\ref{fig_MomentsPhaseVsTime}, where the signal $\left< \Delta \phi_3^2 \right>$ is averaged in the regions "A", "V", and "B", depicted in Fig.~\ref{fig_Images}(e) and plotted versus time. The scaling relations between the reported moments and velocity fields will be the subject of further examination.
%


In conclusion, we demonstrated the measurement of blood flow contrasts in the vascular tree surrounding the optic nerve head in the retina of a pigmented rat by broadband holographic interferometry with a high throughput camera, used at a sampling rate of 39 kHz. In the reported experiment, tissue exposure was limited to 1.6 mW of continuous wave laser radiation at 785 nm over $\sim$ 3 mm $\times$ 3 mm during less than 2 seconds. Interferograms were recorded with an off-axis Mach-Zehnder interferometer with a high throughput camera. Hologram rendering and analysis were performed offline. The three first moments of the envelope of the short-time Fourier transform of holograms enabled robust, high spatial and temporal resolution mapping of Doppler contrasts from flow velocity over a large field of view. Observation of pulsatile retinal blood flow contrasts over 400 $\times$ 400 pixels with a spatial resolution of $\sim 8$ microns and a temporal resolution of 6.5 ms was achieved. In addition, high speed recordings enabled the measurement of optical phase contrasts from which the projection of local flow velocity along the optical axis, and the local variance and pulsatility of the flow were assessed. The first moments of the frequency and phase shifts enabled assessment of flow direction, and the second moments allowed discrimination of its pulsatility in arteries and veins. High-speed near-infrared holography is a robust way of providing a wealth of quantitative information for the noninvasive investigation of local retinal dynamics which could enable novel investigations of blood flow, in particular the analysis of its regional variations.


This work was supported by Fondation Pierre-Gilles de Gennes (FPGG014), the Investments for the Future program (LabEx WIFI: ANR-10-LABX-24, ANR-10-IDEX-0001-02 PSL*), and European Research Council (ERC Synergy HELMHOLTZ, grant agreement \#610110).


\end{document}